\documentstyle[epsfig,amssymb]{aipproc}
\begin{document}

\newcommand{\EP}{\mbox{e$^+$}}
\newcommand{\EM}{\mbox{e$^-$}}
\newcommand{\EPEM}{\mbox{e$^+$e$^-$}}
\newcommand{\EMEM}{\mbox{e$^-$e$^-$}}
\newcommand{\EE}{\mbox{ee}}
\newcommand{\GG}{\mbox{$\gamma\gamma$}}
\newcommand{\GP}{\mbox{$\gamma$e$^+$}}
\newcommand{\GE}{\mbox{$\gamma$e}}
\newcommand{\LGE}{\mbox{$L_{\GE}$}}
\newcommand{\LGG}{\mbox{$L_{\GG}$}}
\newcommand{\LEE}{\mbox{$L_{\EE}$}}
\newcommand{\TEV}{\mbox{TeV}}
\newcommand{\WGG}{\mbox{$W_{\gamma\gamma}$}}
\newcommand{\GEV}{\mbox{GeV}}
\newcommand{\EV}{\mbox{eV}}
\newcommand{\CM}{\mbox{cm}}
\newcommand{\M}{\mbox{m}}
\newcommand{\MM}{\mbox{mm}}
\newcommand{\NM}{\mbox{nm}}
\newcommand{\MKM}{\mbox{$\mu$m}}
\newcommand{\E}{\mbox{$\epsilon$}}
\newcommand{\EN}{\mbox{$\epsilon_n$}}
\newcommand{\EI}{\mbox{$\epsilon_i$}}
\newcommand{\ENI}{\mbox{$\epsilon_{ni}$}}
\newcommand{\ENX}{\mbox{$\epsilon_{nx}$}}
\newcommand{\ENY}{\mbox{$\epsilon_{ny}$}}
\newcommand{\EX}{\mbox{$\epsilon_x$}}
\newcommand{\EY}{\mbox{$\epsilon_y$}}
\newcommand{\SEC}{\mbox{s}}
\newcommand{\CMS}{\mbox{cm$^{-2}$s$^{-1}$}}
\newcommand{\MRAD}{\mbox{mrad}}
\newcommand{\IND}{\hspace*{\parindent}}
\newcommand{\beq}{\begin{equation}}
\newcommand{\eeq}{\end{equation}}
\newcommand{\beqn}{\begin{eqnarray}}
\newcommand{\eeqn}{\end{eqnarray}}
\newcommand{\dst}{\displaystyle}
\newcommand{\bm}{\boldmath}
\newcommand{\BX}{\mbox{$\beta_x$}}
\newcommand{\BY}{\mbox{$\beta_y$}}
\newcommand{\BI}{\mbox{$\beta_i$}}
\newcommand{\SX}{\mbox{$\sigma_x$}}
\newcommand{\SY}{\mbox{$\sigma_y$}}
\newcommand{\SZ}{\mbox{$\sigma_z$}}
\newcommand{\SI}{\mbox{$\sigma_i$}}
\newcommand{\SIP}{\mbox{$\sigma_i^{\prime}$}}
\newcommand{\n}{\mbox{$n_f$}}

\title{Introduction and recent developments in \GG,\GE\ colliders}
\author{Valery Telnov}
\address{ Institute of Nuclear Physics,
630090, Novosibirsk, Russia \\ 
and DESY, Germany} 
\maketitle

\begin{abstract}
 
  High energy photon colliders (\GG, \GE) based on backward Compton
  scattering of laser light is a very natural addition to \EPEM\ 
  linear colliders. In this report we consider mainly this option for
  the TESLA project.  Recent study has shown that the horizontal
  emittance in the TESLA damping ring can be further decreased by a
  factor of four.  In this case the \GG\ luminosity luminosity in the
  high energy part of spectrum can reach 0.3--0.5 $L_{\EPEM\ }$.
  Typical cross sections of interesting processes in \GG\ collisions
  are higher than those in \EPEM collisions by about one order of
  magnitude, so the number of events in \GG\ collisions will be more
  that in \EPEM\ collisions.  The key new element in photon colliders
  is a very powerful laser system. 
  The  most straightforward solution is ``an
  optical storage ring (optical trap)'' with diode pumped laser
  injector which is today technically feasible.  This paper briefly
  review the status of a photon collider based at TESLA, its possible
  parameters.

\end{abstract}

\section{Introduction}

Over the last decade, several laboratories in the world have been
working on linear \EPEM\ collider projects with an energy from several
hundreds GeV up to several TeV. Beside \EPEM\ collisions, linear colliders can
``convert'' electrons to high energy photons using the Compton
backscattering of laser light, thus obtaining \GG\ and $\gamma e$
collisions with energies and luminosities close to those in \EPEM\ 
collisions~\cite{GKST83,TEL90}.  Recently the
International Workshop on High Energy Photon Colliders has been held
at DESY. There are very good news, which I will discuss below.

\begin{figure}[!hbt]
\centering
\vspace*{-1.cm}
\epsfig{file=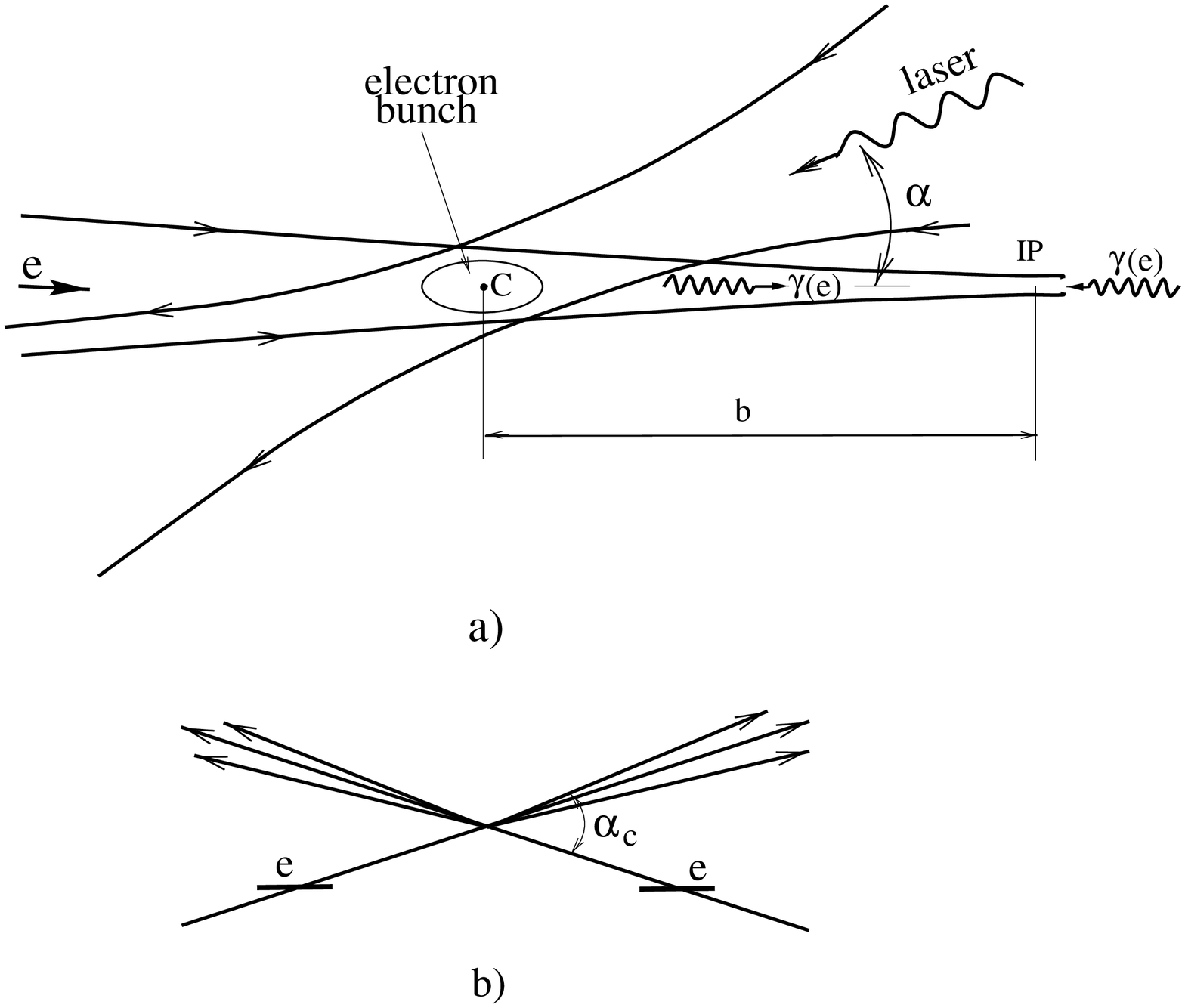,width=8.cm,angle=0}
\vspace*{-0.cm}
\caption{Scheme of  \GG, \GE\ collider.}
\label{ggcol}
\end{figure}

The basic scheme of a photon collider is shown in Fig.~\ref{ggcol}. 

 The maximum energy of the scattered photons (in
direction of electrons) is~\cite{GKST83}
\begin{equation}
\omega_m=\frac{x}{x+1}E_0; \;\;\;\;
x = \frac{4E_0\omega_0\cos^2{\alpha/2}}{m^2c^4}
 \simeq 15.3\left[\frac{E_0}{\TEV}\right]
\left[\frac{\omega_0}{eV}\right],
\end{equation}
where $E_0$ is the electron energy, $\omega_0$ the energy of the laser
photon, $\alpha$ is the angle between electron and laser beam (see,
fig.\ref{ggcol}a). 
For example: $E_0$ =250\,\, GeV, $\omega_0 =1.17$ eV ($\lambda=1.06$
\MKM) (Nd:Glass and other powerful solid state lasers) $\Rightarrow$
$x=4.5$ and $\omega/E_0 = 0.82$.
For an introduction to photon colliders see
refs~\cite{GKST83,TEL90}.

Below I will discuss most important issues connected with a  photon
collider based on the TESLA~\cite{Brinkmann98}.
The following essential topics are discussed: physics motivation,
 parameters of the photon collider at TESLA, lasers, optics.

\section{Physics}

In general, physics in \EPEM\ and \GG, \GE\ collisions 
is quite similar because the same particles can be produced.
However, it is always better to study new phenomena in various
reactions because they give complementary information.  Some
phenomena can best be studied at photon colliders due to
better accuracy or larger accessible masses.

The second aspect important for physics motivation is the luminosity.
Typical luminosity distribution in \GG\ and \GE\ collisions has a high
energy peak and some low energy part (see the next section).  This
peak has a width at half  maximum of about 15\% (in \GG\ collision)
and photons here can have a high degree of polarization.  In the next
section we will see that in the current TESLA designs the \GG\ 
luminosity in the high energy peak  can be
up to 30-40\% of the \EPEM\ luminosity at the same beam energy.

{\bf Higgs boson}

The present Standard Model (SM) assumes existence of a very unique
particle, the Higgs boson. It has not been found yet, but from
existing experimental information it follows that, if it exists, its
mass is higher then 112 GeV (LEP200) and is below
200 GeV, i.e.  lays in the region of the next linear
colliders.  In the simplest extensions to the SM the Higgs sector
consists of five physics states: $h^0,H^0,A^0$ and $H^{\pm}$. All
these particles can be studied at photon colliders and some
characteristics can be measured better than in \EPEM\ collisions.

The process $\gamma\gamma \to H$ goes via the loop with heavy virtual charged
particles and its cross section is very sensitive to the contribution 
of particles with masses far beyond the energies covered by present and planned
accelerators.  
 For the integrated luminosity
50 fb$^{-1}$ (in the peak) the number of produced Higgs will be 50--150
thousands (depending on the mass).
As a result, one can measure the
$\Gamma_{\GG}(H)$ width at photon colliders with an accuracy better
than 2-3\%~\cite{Rembold99}. 
This is sufficient for distinguishing between Higgs models
\cite{GinzKrav}.

Moreover, in the models with several neutral Higgs bosons, heavy H$^0$
and A$^0$ bosons have almost equal masses and at certain parameters in the
theory are produced in \EPEM\ collisions only in associated production
$\EPEM\ \to HA$, while in \GG\ collision they can be
produced singly with sufficiently high cross
section. Correspondingly, in \GG\ collisions one
can produce Higgs bosons with about 1.5 times higher masses.

{\bf Charge pair production}

The second example is the charged pair production. 
 Cross sections for the production
of charged scalar, lepton, WW pairs in \GG\ collisions are larger than
those in \EPEM\ collisions by a factor of approximately 5--20.
The corresponding graphs can be found elsewhere~\cite{TEL90,Tfrei}.

Note, that in \EPEM\ collisions two charged pairs are produced both via
annihilation diagram with virtual $\gamma$, $Z$ and also via
exchange diagrams where some new particles can give contributions,
while in \GG\ collisions it is pure a QED process which allows the
charge of produced particles to be measured unambiguously.  

{\bf Accessible masses}

In \GE\ collisions, charged particle with a mass higher than that in
\EPEM\ collisions can be produced (a heavy charged particle plus a
light neutral); for example, supersymmetric charged particle plus
neutralino or new W boson and neutrino.  Also, \GG\ collisions  provide
higher accessible masses for particles which are produced as a single
resonance in \GG\ collisions (such as neutral Higgs bosons).

{\bf Search for anomalous interactions}

Precise measurement of cross sections allow the observation of effects of
anomalous interactions. The process $\GG\ \to$ WW has large cross
section (about 80 pb) and it is one of most sensitive processes for
a search for a new physics.
The vertex $\gamma WW$ can be studied much better than in \EPEM\ 
collisions because in the latter case the cross section is much smaller
and this vertex gives only 10\% contribution to the total cross
section. The two factors together give about 40 times difference in the cross
sections.  Besides that, in \GG\ collisions the $\gamma\gamma
WW$ vertex can be studied.

{\bf Quantum gravity effects with Extra Dimensions}

This new theory  suggests a possible explanation of why
gravitation forces are so weak in comparison with electroweak forces.
It is suggested that the gravitation constant is equal to the electroweak but
in a space with extra dimensions.  It can be
tested at photon colliders and a two times higher mass scale than in \EPEM\ 
collisions can be reached~\cite{RIZZO}.

Many other examples can be found in proceedings of GG2000
Workshop~\cite{GG2000}.

\section{Possible luminosities of \GG,\GE\ collisions at TESLA}

As it is well known in \EPEM\ collisions the luminosity is restricted
by beamstrahlung and beam instabilities. Due to the first effect the
beams should be very flat. In \GG\ collisions these effects are
absent, therefore one can use beams with much smaller cross section.
At present TESLA beam parameters the \GG\ luminosity is
determined only by the attainable geometric \LEE\ luminosity. 
So, the \GG\ luminosity depends on emittances of electron beams.

Specially for photon collider the TESLA group has studied the
possibility of decreasing emittances at the TESLA damping ring.  The
conclusion is that the horizontal emittance can be reduced by a factor
of 4 in comparison with the previous design.

The resulting parameters of the photon collider at TESLA for 2E=500
GeV and H(130) are presented in Table 1~\cite{GGTEL}. It is assumed
that electron beams have 85\% longitudinal polarization and laser
photons have 100\% circular polarization and that $L \propto E$. The
thickness of laser target is one collision length (so that $k^2
\approx 0.4$).

\begin{table}[!hbtp]
\caption{Parameters of  the \GG\ collider based on TESLA. 
Left column for 2E=500 GeV, next two columns for Higgs with M=130 GeV, 
two options.}
{\renewcommand{\arraystretch}{1.2}
\begin{center}
\begin{tabular}{l c c c} 
 & 2E=500 & 2E=200 & 2E=158 \\
 & $x=4.6$ & $x=1.8$ & $x= 4.6$ \\  \hline 
$N/10^{10}$& 2 & 2 & 2  \\  
$\sigma_{z}$, mm& 0.3 & 0.3 & 0.3  \\  
$f_{rep}\times n_b$, kHz& 14.1 & 14.1 & 14.1  \\
$\gamma \epsilon_{x/y}/10^{-6}$,m$\cdot$rad & 2.5/0.03 & 2.5/0.03 & 
2.5/0.03 \\
$\beta_{x/y}$,mm at IP& 1.5/0.3 & 1.5/0.3 & 1.5/0.3 \\
$\sigma_{x/y}$,nm & 88/4.3 & 140/6.8 & 160/7.6  \\  
\LEE(geom), $10^{33}$& 120 & 48 &  38 \\  
$\LGG (z>0.8z_{m,\GG\ }),10^{33} $ & 11.5 & 3.5 &  3.6  \\
$\LGE (z>0.8z_{m,\GE\ }),10^{33}$ & 9.7 & 3.1 & 2.7 \\
\end{tabular}
\end{center}
}
\label{table1}
\end{table}   

\vspace{-3mm}

For these luminosities the rate of production of the SM Higgs boson
with M$_H$=130(160) GeV in \GG\ collisions is 0.9(3) of that in \EPEM\ 
collisions at 2E = 500 GeV (both reactions, ZH and H$\nu\nu$).

Comparing the \GG\ luminosity with the \EPEM\ luminosity
($L_{\EPEM} = 3\times 10^{34}$ \CMS\ for $2E=500$ GeV)
we see that for the same energy
\begin{equation}
\LGG(z>0.8z_m) \sim 0.4 L_{\EPEM}. 
\label{lgge+e-}
\end{equation}
For example, the cross section for
production of $H^+H^-$ pairs in collisions of polarized photons is higher
than that in \EPEM\ collisions by a factor of 20 (not far from the
threshold); this means 8 times higher production rate for the
luminosities given above.
The relation (\ref{lgge+e-}) is valid only for the considered beam
parameters. A more universal relation is (for $k^2=0.4$)
\begin{equation}
\LGG(z>0.8z_m) \sim 0.1\LEE(geom).   
\label{lgge-e-}
\end{equation}

Figures for the luminosity distribution in \GG\ and \GE\ collisions
can be found elsewhere~\cite{GGTEL}.

\section{Lasers, optics}

A key element of photon colliders is a powerful laser system which is
used for e$ \to\gamma$ conversion.  Lasers with the required flash
energies (several J) and pulse duration $\sim$ 1 ps already exist and
are used in several laboratories. The main problem is the high
repetition rate, about 10--15 kHz, with a pulse structure repeating
the time structure of the electron bunches.  

\begin{figure}[!htb]
\centering
\vspace*{0.cm} 
\hspace*{-0.4cm} \epsfig{file=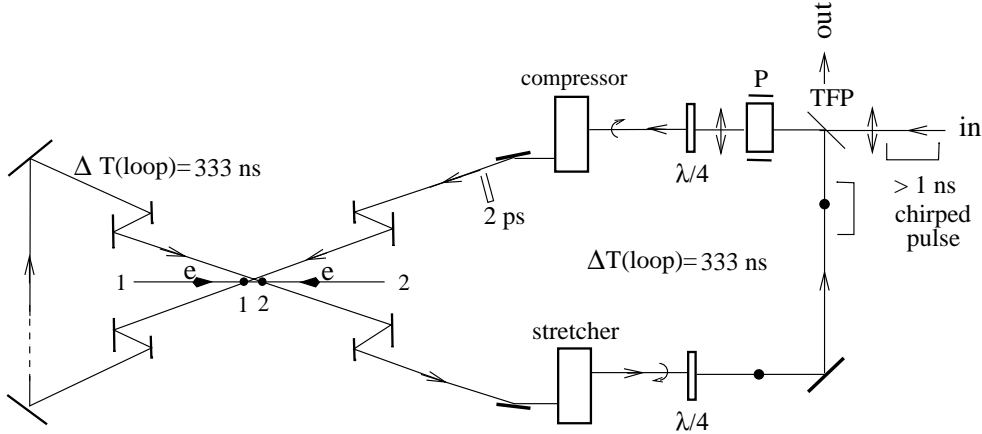,width=13cm,angle=0} 
\vspace*{0.2cm} 
\caption{Optical storage ring for $e \to \gamma$ conversions. 
P is a Pockels cell, TFP is a thin film polarizer, thick dots and 
double arrows show the direction of polarization.}
\vspace{0mm}
\label{loop}
\vspace{-0mm}
\end{figure}

The most attractive and reliable solution at this moment is
an ``optical storage ring'' with a diode pump laser injector.
 This new approach can be considered as
a base-line solution for the TESLA photon collider~\cite{GGTEL}.
This scheme with multiple use of each laser bunch is shown in fig.~\ref{loop}.

The laser pulses are send to the interaction region
where they are trapped in an optical storage ring. This can be done using
Pockels cells (P), thin film polarizers (TFP) and 1/4-wavelength
plates ($\lambda/4$). Each bunch makes several (n) round trips (I assume
n=10 for further example) and then is deleted from the ring. All these
tricks can be done by switching one Pockels cell.

During one total loop each bunch is used for conversion twice.
To avoid problems of non-linear effects (self-focusing) in optical
elements, laser pulses which are compressed before collisions down to
about 2 ps using grating pairs are then stretched again up to previous
length before passing through optical elements.  Such system can be
done in the best way only at the TESLA due to a large spacing between
bunches and a long train.

 A laser system required for a such optical storage
ring with 10 round trips can consist of about 8 lasers of 0.9 kW
average powereach. 
Due to the high average power the lasers should be based on diode
pumping. Diodes have much higher efficiency than flash lamps and much
more reliable. This technology is developed very actively for other
application, such as inertial fusion.
Present cost of diodes for such laser system is about 15 M\$ (for
10-fold use of one laser bunch as described above) and it is expected that
their cost will be further decreased several times.
  Such system can be done now:  all technologies exist.

\section{Conclusion}

The luminosity in \GG\ collisions (in the high
energy peak) can reach about 40\% of \EPEM\ luminosity. Since cross
sections in \GG\ collisions are typically higher by one order of
magnitude than those in \EPEM\ collisions and because of 
access to higher masses for some particles, the photon
collider now has very serious physics motivation.

There is good suggestion for the laser system, which, it seems, can be
build now.


\begin{thebibliography}{99}
%

\bibitem{GKST83} I.Ginzburg, G.Kotkin, V.Serbo, V.Telnov,{\it Nucl.Instr.
{\rm\&} Meth.} {\bf 205} (1983) 47 (Prepr. INP 81-102, Novosibirsk, 1991).



\bibitem{TEL90} V.Telnov, {\it Nucl.Instr.{\rm\&}  Meth.} {\bf A 294}
  (1990) 72;{\it Nucl.Instr.{\rm\&}Meth.} {\bf A 355} (1995) 3.


\bibitem{Brinkmann98} R.Brinkmann et al., {\it Nucl. Instr.  {\rm\&}Meth. }
  {\bf A 406} (1998) 13.


\bibitem{Tfrei} V.~Telnov, Proc.  of the International Conference on the 
Structure and Interactions of the Photon (Photon 99), Freiburg,
Germany, 23-27 May 1999,  to be published in Nucl. Phys. Proc. Suppl. B, 
e-print: hep-ex/9908005. 

\bibitem{Rembold99} G.Jikia, S.Soldner-Rembold, Proceedings of 4th
  International Workshop on Linear Colliders (LCWS 99), Sitges,
  Barcelona, Spain, 28 Apr - 5 May 1999.  e-print: hep-ph/9910366.


\bibitem{GinzKrav} I.Ginzburg, M.Krawczyk,  Proceedings of 4th
  International Workshop on Linear Colliders (LCWS 99), Sitges,
  Barcelona, Spain, 28 Apr - 5 May 1999, hep-ph/9909455 and these proceedings.

 
\bibitem{GG2000} Proc. of Intern. Workshop on High Energy Photon
  Colliders, 14-17 June, 2000, DESY, Hamburg, Germany, to be published
  in Nucl. Inst. and Methods A.
%
\bibitem{RIZZO} Thomas Rizzo, {\it Proceedings of 4th International
  Workshop on Linear Colliders (LCWS 99)}, Sitges, Barcelona, Spain, 28
  Apr - 5 May 1999, SLAC-PUB-8204, e-Print Archive: hep-ph/9907401, also
  these proceedings. 

\bibitem{GGTEL} V.Telnov, Proc. of Intern. Workshop on High Energy Photon
  Colliders, 14-17 June, 2000, DESY, Hamburg, Germany, to be published
  in Nucl. Inst. and Methods A.




\end{thebibliography}
\end{document}